\documentclass[a4paper,nojss]{jss}
\usepackage{amsmath,amssymb,amsfonts,thumbpdf, caption, subcaption, lipsum, hyperref}

\DefineVerbatimEnvironment{example}{Verbatim}{}
\setkeys{Gin}{width=\textwidth}

\newcommand*{\ditto}{--\,\raisebox{-0.5ex}{\texttt{"}}\,--}

\usepackage{chngcntr}

\shortcites{kunkel.etal:2012,vose.etal:2013}

\title{Bivariate Gaussian models for wind vectors in a distributional regression framework}
\Plaintitle{Bivariate Gaussian models for wind vectors}

\author{Moritz N. Lang\\Universit\"at Innsbruck \And
        Georg J. Mayr\\Universit\"at Innsbruck \And
        Reto Stauffer\\Universit\"at Innsbruck \And
        Achim Zeileis\\Universit\"at Innsbruck}
\Plainauthor{Moritz N. Lang, Georg J. Mayr, Reto Stauffer, Achim Zeileis}

\Abstract{
A new probabilistic post-processing method for wind vectors is presented in a
distributional regression framework employing the bivariate Gaussian
distribution. In contrast to previous studies all parameters of the
distribution are simultaneously modeled, namely the means and variances for
both wind components and also the correlation coefficient between them
employing flexible regression splines. To capture a possible mismatch between
the predicted and observed wind direction, ensemble forecasts of both wind
components are included using flexible two-dimensional smooth functions. This
encompasses a smooth rotation of the wind direction conditional on the season
and the forecasted ensemble wind direction. 

The performance of the new method is tested for stations located in plains,
mountain foreland, and within an alpine valley employing ECMWF ensemble
forecasts as explanatory variables for all distribution parameters. The
rotation-allowing model shows distinct improvements in terms of predictive
skill for all sites compared to a baseline model that post-processes each wind
component separately. Moreover, different correlation specifications are tested
and small improvements compared to the model setup with no estimated
correlation could be found for stations located in alpine valleys.}

\Plainkeywords{bivariate Gaussian distribution, distributional regression, regression splines, wind vectors, ensemble forecasts}

\Address{
  Moritz N. Lang\\
  Department of Statistics\\
  Department of Atmospheric and Cryospheric Sciences\\
  Universit\"at Innsbruck\\
  6020 Innsbruck, Austria\\
  E-mail: \email{moritz.lang@uibk.ac.at}\\

}

\begin{document}

\section{Introduction}\label{sec:intro}
Accurate forecasts of wind speed and direction are of great importance for
decision making processes and risk management in today's society and will
likely become more important in the future. Not only because of the rapid
change in climate and the resulting increase of severe storms
\citep[e.g.,][]{kunkel.etal:2012, vose.etal:2013}, but also due to the change
of the society itself and its technical revolution. As an example, the European
Union is aiming to increase the amount of wind energy by 2030 to 35\%, which
would be more than double the capacity installed at the end of 2016
\citep{windeurope:2017}. In the field of aviation and air traffic control for
instance, more flexible landing procedures with a so-called time-based
separation are currently tested at Heathrow Airport and are planned to go
operational in the near future \citep{europeancommission:2018}. In both cases,
wind (power) forecasts are of fundamental importance; probabilistic wind
forecasts are in particular advisable as they permit optimal risk assessment
and decision making \citep{gneiting:2008}. 

Probabilistic weather forecasts are usually issued in form of ensemble
predictions. To account for the underlying uncertainty in the atmosphere,
numerical ensemble prediction systems (EPSs) provide a set of weather forecasts
using slightly perturbed initial conditions and different model
parameterizations \citep{palmer:2002}. Despite recent advances in the
development of EPSs, the resulting forecasts still often show displacement
errors and usually capture only part of the forecast uncertainty, especially
when comparing EPS forecasts and point measurements \citep{buizza.etal:2005,
gneiting.katzfuss:2014}. This often results from structural model deficiencies,
and insufficient resolution or unresolved topographical features. To remove
systematic biases and to provide corrected variance information, statistical
post-processing methods are often employed. For wind, various ensemble
post-processing methods have been proposed over the last decade, mainly
focusing on wind speed. For a single location, parametric examples are
non-homogeneous regression \citep{thorarinsdottir.gneiting:2010,
lerch.thorarinsdottir:2013, baran.lerch:2015, baran.lerch:2016}, kernel
dressing methods with similarities to Bayesian model averaging
\citep{sloughter.etal:2010, courtney.etal:2013, baran:2014}, and extended
logistic regression \citep{messner.etal:2014, messner.etal:2014c}. A
non-parametric approach based on quantile regression forests was applied by
\citet{taillardat.etal:2016}. On a regular grid, ensemble post-processing based
on non-homogeneous regression was performed by \citet{scheuerer.moller:2015}.

To account for the circular characteristics of wind or utilizing information of
wind speed and direction, an intuitive post-processing approach is to model a
bivariate process for the zonal and meridional wind components.
\citet{gneiting.etal:2008} suggested to use a bivariate Gaussian response
distribution for the wind components, which \citet{pinson:2012} implemented. He
estimates a dilation and translation factor for the individual ensemble members
utilizing the empirical correlation structure of the EPS. This procedure can be
seen as a variant of the ensemble copula coupling (ECC) method introduced by
\citet{schefzik.etal:2013}. With the ECC, both wind components are calibrated
with univariate approaches and a discrete sample drawn from each univariate
predictive distribution is rearranged in the rank order structure of the raw
ensemble. The method introduced by \citet{schuhen.etal:2012} also fits a
bivariate Gaussian distribution for the wind components; however, in their
approach the post-processed probabilistic forecast consists of a fully
specified predictive distribution instead of a discrete ensemble. As their
analyses show that the correlation mainly depends on wind direction, they model
the correlation as a trigonometric function of the ensemble mean wind
direction, separately for different wind sectors. In addition, an extra group
is formed for cases with low wind speeds unconditionally on their wind
direction. The estimation of the correlation parameter is done offline in a
pre-processing step for a separate year either for all stations combined or for
each station individually. However according to \citet{schuhen.etal:2012}, the
fitting can be critical for individual stations since wind sectors may only
contain few data points.

For stations in complex terrain, a possible drawback of the bivariate
post-processing approaches introduced by \citet{pinson:2012} and
\citet{schuhen.etal:2012} is that the models are not able to correct for a
systematic distortion in the wind directions due to discrepancies between the
model and real topography. Especially, when the respective valley orientations
differ a meridional wind component might be partially rotated into a zonal wind
component and vice versa. In the field of post-processing deterministic weather
forecasts, \citet{glahn.lowry:1972} already suggested to employ both forecasted
wind components for the calibration of the zonal or meridional wind component,
which is partly able to correct for systematic distortions in wind directions
in a linear manner. However, to our knowledge, this approach has not yet been
further pursued in the field of bivariate probabilistic post-processing of wind
vectors.

Alternatively to bivariate calibration methods, wind direction can also be
employed in univariate settings. In a post-processing approach for wind speed,
\citet{eide.etal:2017} suggest to utilize the potentially non-linear
information of the wind direction by a generalized additive
model~\citep[GAM;][]{hastie.tibshirani:1986}. GAMs were first applied in the
meteorological context by \citet{vislocky.fritsch:1995}, and provide a powerful
statistical model framework which can capture potential non-linear
relationships between the covariates and the response by smooth functions or
splines. \citet{eide.etal:2017} employ wind direction as an additional
covariate for the estimation of wind speed, by accounting for its cyclic and
potential non-linear characteristics utilizing thin-plate regression splines. 

In this study, we directly model the zonal and meridional wind components
employing the bivariate Gaussian distribution as suggested by
\citet{gneiting.etal:2008}, and performed by \citet{pinson:2012} and
\citet{schuhen.etal:2012}. However, we capture all distribution
parameters, namely the location and scale parameters for both wind components, and also
the correlation coefficient between them in a single flexible model. In the estimation of the
two-dimensional location and scale parameters the information value of both
ensemble wind components is utilized to allow for a smooth rotation of the
forecasted wind direction accounting for unresolved topographical features. To
consider the correlation characteristics detected in \citet{schuhen.etal:2012}
and to allow for possible non-linear effects, as e.g., suggested by
\citet{eide.etal:2017}, we model the correlation as a function of wind speed
and direction utilizing cyclic regression splines. To account for potential
time-varying effects, all linear predictors use a time-adaptive intercept and
time-adaptive slope coefficients based on cyclic smooth splines.

The paper is structured as follows: Section~\ref{sec:method} introduces the
employed statistical models. The underlying data of this study are shortly
described in Sect.~\ref{sec:data}. Within Sect.~\ref{sec:results}, first a
model comparison and validation is presented for two weather stations with
different site characteristics, followed by aggregated scores for station sites
located in plains, mountain foreland, and within an alpine valley. The article
ends with a brief discussion and a conclusion given in Sect.~\ref{sec:disc}.

\section{Methods}\label{sec:method}
In Sect.~\ref{sec:method:biv}, the bivariate Gaussian distribution is reviewed
and briefly presented in a distributional regression framework. Subsequently,
three broad model classes are introduced all of which are based on a
time-adaptive training scheme but employ different specifications for the
location, scale, and correlation parameters of the bivariate distribution.
First, the baseline model is presented in Sect.~\ref{sec:method:blm-0} that
serves as a benchmark and simply combines two univariate heteroscedastic
regression models that post-process each wind component separately. Second, the
baseline model is extended in Sect.~\ref{sec:method:ram-0} by always adding
both EPS wind components as regressors using smooth splines and thus allowing
for potential misspecifications in the EPS wind direction. Finally,
Sect.~\ref{sec:method:ram-x} also considers models with estimated correlation
coefficients based on various regression specifications.
Table~\ref{tab:models} provides a synoptic summary of all bivariate Gaussian
model specifications tested within this study.
 
\subsection{Distributional regression for a bivariate Gaussian response}\label{sec:method:biv}
The zonal and meridional components of the horizontal wind vector are
represented by a bivariate Gaussian distribution. Its likelihood function $L$
is given by

\begin{equation}
   L(\boldsymbol\mu, \Sigma | \boldsymbol y) = \frac{1}{\sqrt{(2\pi)^2|\boldsymbol\Sigma|}}
   \exp\left(-\frac{1}{2}({\boldsymbol y}-{\boldsymbol\mu})^\top{\boldsymbol\Sigma}^{-1}
   ({\boldsymbol y}-{\boldsymbol\mu})
   \right),\label{eq:bivnorm}
\end{equation}

where $\boldsymbol y = (y_1, y_2)^\top$ are bivariate observations and
$\boldsymbol\mu = (\mu_1, \mu_2)^\top$ the distributional location parameters
with $\mu_\star \in \mathbb{R}$; the subscript asterisk acts as a placeholder
for the zonal and meridional wind component from here on. The covariance matrix
is defined as

\begin{equation}
  \boldsymbol\Sigma = \begin{pmatrix} \sigma_1^2 & \rho \sigma_1 \sigma_2 \\ 
    \rho \sigma_1 \sigma_2  & \sigma_2^2 \end{pmatrix},
\end{equation}

with correlation parameter $\rho \in [-1, 1]$ and scale parameters
$\sigma_{\star} > 0$. In the framework of distributional regression, the
location parameters $\mu_{\star}$, the scale parameters $\sigma_{\star}$, and
the correlation parameter $\rho$ are linked to additive predictors by an
identity, logarithm and rhogit link, respectively \citep{klein.etal:2014}. 

To be able to utilize the information of cyclic covariates as e.g., wind
direction in addition to linear covariates, we follow \citet{eide.etal:2017}
and fit a GAM but utilize cubic smooth functions with cyclic constraints for
all cyclic covariates (\citealt{wood:2017}; see Appendix~\ref{app:gam}). In the
context of distributional regression, additive models with smooth effects are
typical referred to as `generalized additive models for location, scale and
shape'~\citep[GAMLSS, ][]{rigby.stasinopoulos:2005}. In this study, we utilize
GAMLSS in a Bayesian framework, which allows us to examine the estimated
effects based on Markov chain Monte Carlo (MCMC) simulations. A comprehensive
summary of the method is given in \citet{umlauf.etal:2018}. 

\subsection{Baseline model (BLM-0)}\label{sec:method:blm-0}
The baseline model (\mbox{BLM-0}) combines two univariate heteroscedastic
regression models that post-process each wind component separately with correlation
fixed at zero. Hence, for the location and scale part, it uses its direct counterparts
of the EPS as covariates, namely {EPS-forecasted} zonal wind
information~($\text{vec}_1$) to model the zonal component of the bivariate
response, and {EPS-forecasted} meridional wind information~($\text{vec}_2$)
to model the meridional component:

\begin{align} 
  \begin{split}
    \mu_{\star} &= 
      \underbrace{\alpha_0 + f_0(\text{doy})}_\text{intercept} + 
      \underbrace{(\alpha_1 + f_1(\text{doy}))}_\text{slope coefficient} 
      \cdot \text{vec}_{\star, mean},\\[1ex]
    \text{log}(\sigma_{\star}) &= 
      \underbrace{\beta_0 + g_0(\text{doy})}_\text{intercept} + 
      \underbrace{(\beta_1 + g_1(\text{doy}))}_\text{slope coefficient} 
      \cdot \text{vec}_{\star, log.sd},\label{eq:gam:locsc:v1}
  \end{split}
\end{align}

where, $\alpha_\bullet$ and $\beta_\bullet$ are regression coefficients, and
$f_\bullet(\text{doy})$ and $g_\bullet(\text{doy})$ employ cyclic
regression splines conditional on the day of the year~(doy). The subscripts
$mean$ and $log.sd$ refer to mean and log standard deviation of the ensemble
wind components, respectively. We follow \citet{gebetsberger.etal:2017} and use
the logarithm transformation for the standard deviation of the ensemble members
to assure positivity which is preferable for the estimation process.

Equation~(\ref{eq:gam:locsc:v1}) specifies a time-adaptive training scheme
(with further details in Appendix~\ref{sec:method:tscheme}), where the linear
predictors consist of a global intercept and slope coefficient plus a
seasonally varying deviation. Thus, the intercept and slope coefficients can
smoothly evolve over the year in case that the bias or the covariate's skill
varies seasonally. If there is no seasonal variation, the non-linear effects
become zero and Eq.~(\ref{eq:gam:locsc:v1}) simplifies to a regression model
with a constant intercept and slope coefficient ($\mu_{\star} = \alpha_0 +
\alpha_1 \cdot \text{vec}_{\star, mean}; ~ \text{log}(\sigma_{\star}) =
\beta_0 + \beta_1 \cdot \text{vec}_{\star, log.sd}$).

\subsection{Rotation-allowing model without correlation ({RAM-0})}\label{sec:method:ram-0}
In the second model, labeled as rotation-allowing model~(\mbox{RAM-0}), we
extend the setup \mbox{BLM-0} by employing the zonal and meridional wind
information of the ensemble for the linear predictors of all location and scale
parameters. That means we use the ensemble information of both the zonal and
meridional wind components for the two components of the response
\citep[cf.][]{glahn.lowry:1972}. In case of a perfect EPS the zonal wind
predictions are non-informative covariates for the meridional wind component
and vice versa. However, if, e.g., the model topography is not sufficiently
resolved or in case of local shadowing effects, both EPS wind components may
contain valuable information for the zonal and meridional wind components of
the response. Especially in a mountain valley, when the model and real valley
orientation differs, both wind components of the ensemble can potentially
contain information about both location and scale parameters, respectively.
Thus, we propose to employ seasonally varying effects depending on the ensemble
wind direction, which allows the model to rotate the forecasted wind direction
if necessary. To do so, we obtain a two-dimensional smooth function with a
respective cyclic constraint for the day of the year~(doy) and for the mean
ensemble wind direction~($\text{dir}_{mean}$):

\begin{equation}
  \begin{split} 
    \mu_{\star} = \alpha_0 + f_0(\text{doy}) &+  
      (\alpha_1 + f_1(\text{doy}) \cdot f_2(\text{dir}_{mean})) \cdot \text{vec}_{1, mean}\\
      &+ (\alpha_2 + f_3(\text{doy}) \cdot f_4(\text{dir}_{mean})) \cdot \text{vec}_{2, mean},\\[1ex]
    \text{log}(\sigma_{\star}) = \beta_0 + g_0(\text{doy}) &+  
      (\beta_1 + g_1(\text{doy}) \cdot g_2(\text{dir}_{mean})) \cdot \text{vec}_{1, log.sd}\\
      &+ (\beta_2 + g_3(\text{doy}) \cdot g_4(\text{dir}_{mean})) \cdot \text{vec}_{2, log.sd},
    \label{eq:gam:locsc:v2}
  \end{split}
\end{equation}

where, as before, $\alpha_\bullet$ and $\beta_\bullet$ are regression
coefficients, and $f_\bullet$ and $g_\bullet$ employ cyclic regression splines.
From a more physical perspective, the two-dimensional smooth effects rotate the
ensemble wind components conditional on the day of the year and the ensemble
wind direction. 

\subsection{Rotation-allowing models with correlation}\label{sec:method:ram-x}
By explicitly modeling the correlation, we further extend the \mbox{RAM-0}
setup within this section. For the estimation of the correlation structure
different model specifications are tested. The most advanced specification,
\mbox{RAM-ADV}, assumes that the correlation mainly depends on the mean
ensemble wind direction~($\text{dir}_{mean}$) and
speed~($\text{spd}_{mean}$) by modeling a linear interaction between these
two covariates:

\begin{align} 
\text{rhogit}(\rho) &= \gamma_0 + h_0(\text{doy}) + 
  h_1(\text{dir}_{mean}) + 
  (\gamma_1 + h_2(\text{dir}_{mean})) \cdot \text{spd}_{mean}\label{eq:gam:rho},
\end{align}

with $\text{rhogit}(\rho) = \rho / \sqrt{(1 - \rho^2)}$; $\gamma_0$ is the
global intercept and $h_0(\text{doy})$ the seasonally varying intercept. The
effect $h_1(\text{dir}_{mean})$ estimates the dependence of the correlation
given the wind direction and $(\gamma_1 + h_2(\text{dir}_{mean})) \cdot
\text{spd}_{mean}$ employs a varying effect of wind speed conditional on the
wind direction. The estimation of the underlying correlation structure is in
accordance to results of \citet{schuhen.etal:2012}, who employ wind direction
and an offset of wind speed as informative covariates in the estimation of the
correlation parameter.

Other implementations tested for the correlation parameter are an
intercept-only model (\mbox{RAM-IC}), a model with a cyclic effect solely
depending on wind direction (\mbox{RAM-DIR}), the univariate model \mbox{RAM-0}
of Sect.~\ref{sec:method:ram-0}, and a univariate model using the empirical
correlation ($corr$) of the raw ensemble (\mbox{RAM-EMP}). A synoptic table of
all models tested in this study is given in Table~\ref{tab:models}.

\begin{table}[tbp!]
  \caption{Overview of bivariate Gaussian model specifications. For the `baseline model'
    (\mbox{BLM-0}, see Sect.~\ref{sec:method:blm-0}) and the `rotation-allowing
     model' (\mbox{RAM-0}, see Sect.~\ref{sec:method:ram-0}) no correlation is
     employed, i.e., fixed at zero.  For all tested correlation specifications, the
     \mbox{RAM-0} setup is employed for the location and scale part
     (see~Sect.~\ref{sec:method:ram-x}). In all setups for each distribution
     parameter, a seasonally varying intercept effect is estimated. For the
     \mbox{BLM-0}, a seasonally varying slope coefficient is fitted for the two wind
     components in the location and scale parts. For the setup \mbox{RAM-0}, the
     slope coefficients are additionally dependent on the wind direction. In the
     correlation model \mbox{RAM-ADV}, the wind speed is modeled conditional on the
     wind direction. The ditto symbol `\ditto' implies that the same configuration
     as in the line above is employed.}
  \begin{center}
    \begin{tabular}{ l c c l}
      \hline
      Name & Location part &  Scale part & Correlation part\\
      \hline
      {BLM-0}   
        & $\mu_{\star} \sim \text{vec}_{\star, mean}$ 
        & $\sigma_{\star} \sim \text{vec}_{\star, log.sd}$ 
        & $\rho = 0$\\
      {RAM-0}   
        & $\mu_{\star} \sim \text{vec}_{1, mean}, \text{vec}_{2, mean}$ 
        & $\sigma_{\star} \sim \text{vec}_{1, log.sd}, \text{vec}_{2, log.sd}$ 
        & $\rho = 0$\\
      {RAM-EMP} 
        & \ditto
        & \ditto
        & $\rho = \text{vec}_{\star, corr}$ \\
      {RAM-IC}
        & \ditto
        & \ditto
        & $\rho \sim 1$\\
      {RAM-DIR}
        & \ditto
        & \ditto
        & $\rho \sim \text{dir}_{mean}$\\
      {RAM-ADV}
        & \ditto
        & \ditto
        & $\rho \sim \text{dir}_{mean},\text{spd}_{mean}$\\
      \hline
    \end{tabular}
    \label{tab:models}
  \end{center}
\end{table}

\section{Data}\label{sec:data}

\subsection{Observational data} 

\begin{figure}[!tb] 
\centering 
\includegraphics[width=0.6\textwidth]{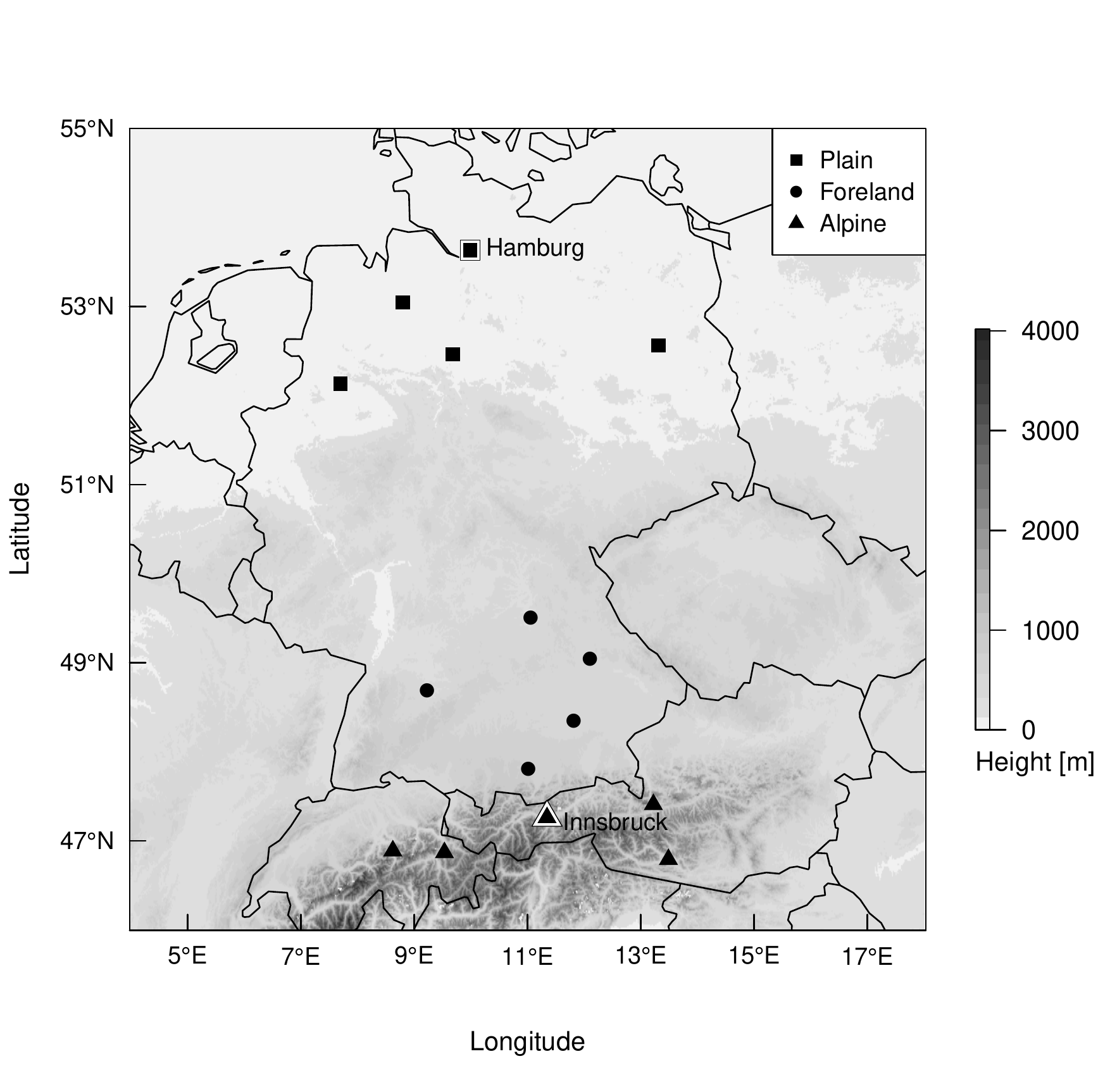}
\caption{Overview of the study area with selected stations classified as plain,
foreland, and alpine station sites. The labeled stations with a white background, 
Hamburg and Innsbruck, are discussed in detail in Sect.~\ref{sec:results}.}
\label{fig:overview} 
\end{figure}

The validation and comparison of the different model specifications is
performed for $15$ measurement sites located across Austria, Germany, and
Switzerland. The sites are chosen to investigate the influence of different
underlying topographies or varying discrepancies between the real and the model
topography on the post-processing. The stations are divided into three groups
representing sites located in plains, mountain foreland, and within an alpine
valley. An overview of the stations is given in Fig.~\ref{fig:overview}. The
results for the stations Hamburg and Innsbruck, which are labeled in
Fig.~\ref{fig:overview}, are discussed in more detail in
Sect.~\ref{sec:results}. At all meteorological sites, wind speed and direction
measurements are reported for the $10$\,m height level. The data are
$10$-minute averages for the period January 26, 2010 to March 7, 2016, yielding
a total of 2233 days.

\subsection{Ensemble prediction system}

\begin{figure}[!tb] 
\centering 
\includegraphics[width=0.8\textwidth]{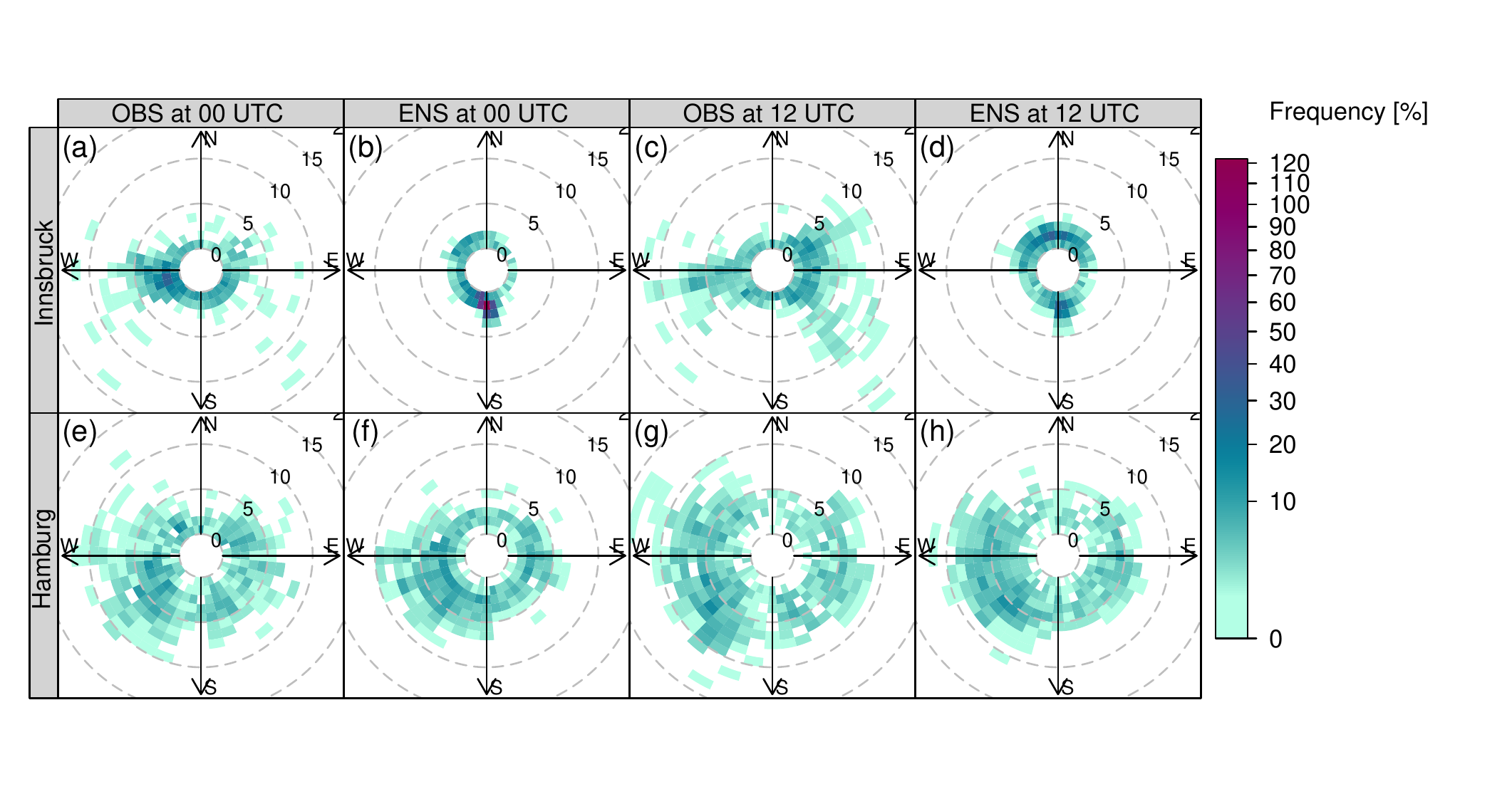}
\caption{Empirical wind distributions of observations~(OBS) and mean ensemble
forecasts~(ENS) for Innsbruck and Hamburg. The probability of occurrence is
color-coded, the wind speed is represented by contour lines (ms$^{-1}$). The
forecast steps $+12$\,h and $+24$\,h, valid at $12$\,UTC and $00$\,UTC, are
shown for the validation period from February 24, 2014 to March 7, 2016.}
\label{fig:climatology} 
\end{figure}

Covariates are derived from the global 50-member EPS of the European Centre for
Medium-Range Weather Forecasts (ECMWF). These EPS forecasts have a horizontal
resolution of approximately $30$\,km (T639) for the time between January 2010
and March 2016 and are bilinearly interpolated to the measurement sites.
Covariates employed in this study are the zonal and meridional wind components
as well as the derived quantities wind speed and direction valid at $10$\,m
above ground. For all these variables, two statistics are computed over the 50
perturbed ensemble members, namely the $mean$ and the logarithm of the standard
deviation ($log.sd$). Additionally, the empirical correlation ($corr$) is
computed from the raw ensemble members to capture their rank dependence
structure. Forecasts are taken from the EPS run initialized at $00$\,UTC for
forecast steps ranging from $+12$\,h to $+72$\,h ahead on a $12$~hourly
temporal resolution. Figure~\ref{fig:climatology} shows the empirical wind
distributions of the observed and predicted winds for Innsbruck and Hamburg for
forecast steps $+12$\,h and $+24$\,h corresponding to $12$\,UTC and $00$\,UTC.

\section{Results}\label{sec:results}
This section presents the results of the statistical post-processing models.
The structure is as follows: First, the estimated effects of the baseline
model~\mbox{BLM-0} (Sect.~\ref{sec:results:blm-0}) and the rotation-allowing
model~\mbox{RAM-0} (Sect.~\ref{sec:results:ram-0}) are shown for two stations
representative for one alpine valley site and for a measurement site in the
plains. For both models a constant correlation of zero is employed, and their
predictive performance is discussed in Sect.~\ref{sec:results:skill-0}.
Afterwards, a model comparison (Sect.~\ref{sec:results:ram-x}) and validation
(Sect.~\ref{sec:results:skill-x}) of the different correlation specifications
is given for the two representative stations. In
Sect.~\ref{sec:results:comparison}, the overall performance of the model setups
is evaluated for three groups of stations classified as topographically plain,
mountain foreland, and alpine valley sites.

The model estimation is performed on data of the first $4$~years, leaving an
out-of-sample validation data set ranging from February 24, 2014 to March 7,
2016.

\subsection{Baseline model ({BLM-0})}\label{sec:results:blm-0}

\begin{figure}[!tb] 
\centering 
\includegraphics[width=0.8\textwidth]{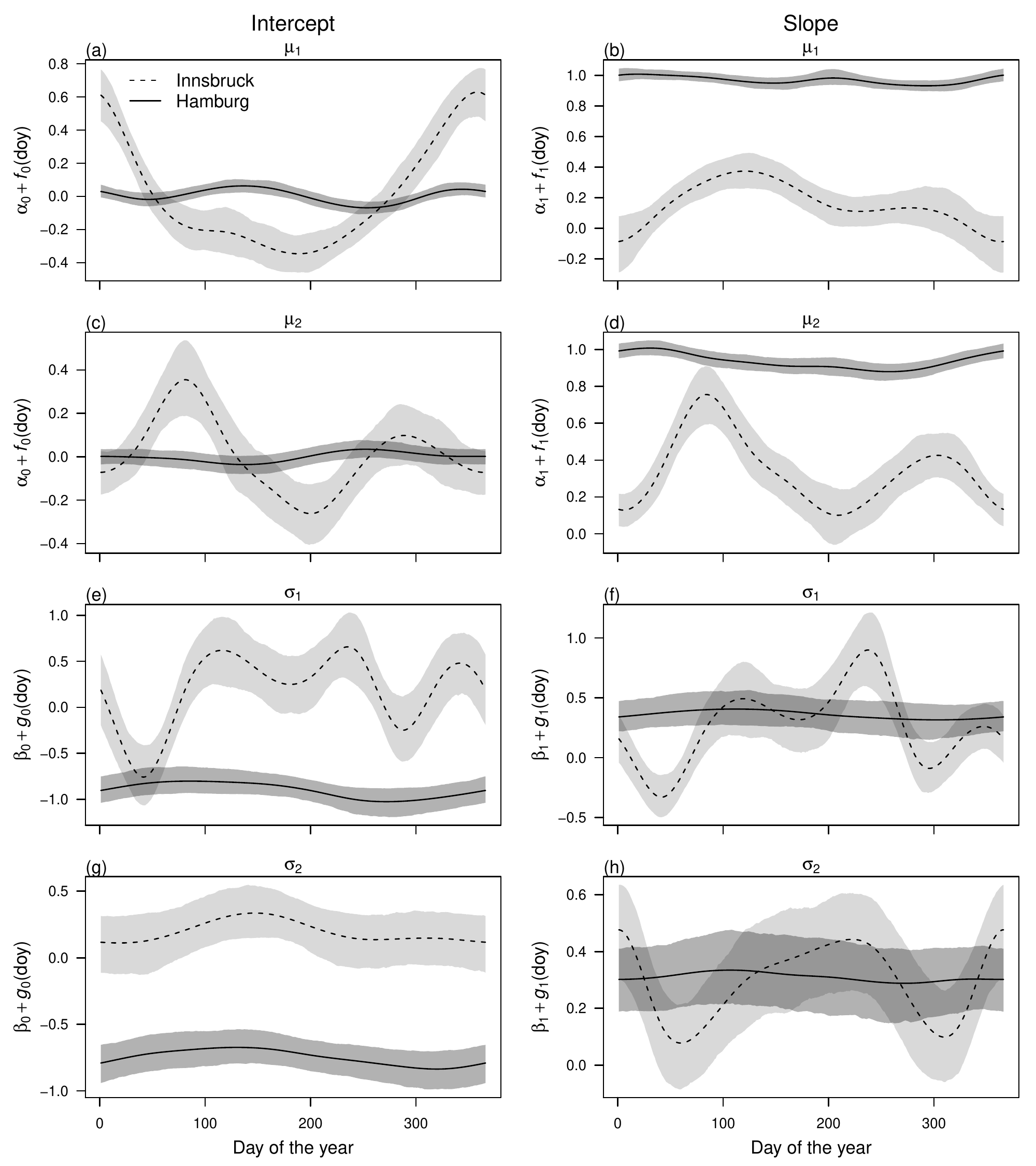}
\caption{
Cyclic seasonal intercept and slope effects according to
Eq.~(\ref{eq:gam:locsc:v1}) employing a constant correlation of zero for the
weather station Innsbruck (dashed) and Hamburg (solid) at forecast step
$+12$\,h (valid at $12$\,UTC). The effects for the location
parameters~$\mu_\star$ (a--d) and scale parameters~$\sigma_\star$ (e--h) are
shown on the linear and log-scale, respectively. The shading represents the
$95$\% credible intervals based on MCMC-sampling.}
\label{fig:effects} 
\end{figure}

For \mbox{BLM-0}, the cyclic seasonal effects for the stations Hamburg and
Innsbruck are shown in Fig.~\ref{fig:effects} as solid and dashed lines
with the respective $95$\% credible intervals. The estimated effects are on the
scale of the additive predictor; i.e., on the linear scale for the location
parameters~$\mu_\star$ and on the log-scale for the scale parameters
$\sigma_\star$. Each of the four distribution parameters is described by a (potentially)
seasonally varying effect for the intercept (left column) and the slope
coefficient (right column) as specified in Eq.~(\ref{eq:gam:locsc:v1}). 

For Hamburg, for both location parameters~$\mu_\star$, the intercept effect is
almost zero (Fig.~\ref{fig:effects}a,\,c) and the effect for the slope
coefficient is close to one (Fig.~\ref{fig:effects}b,\,d) with very
little seasonal variability. This means apparently no bias correction is
necessary and the ensemble mean wind components are mapped almost one-to-one to
the location parameters. Similarly, barely any seasonal variation exists for
the scale parameters~$\sigma_\star$ (Fig.~\ref{fig:effects}e--h);
however, here the intercept and slope coefficients actually post-process the
EPS variances of the wind components (rather than a one-to-one mapping only),
leading to an increase of the scale parameters compared to the under-dispersed
ensemble. The $95$\% credible intervals indicate a higher uncertainty of the
estimated scale parameters compared to the location parameters. In summary, the
EPS performance for Hamburg is almost constant over the year and no
time-adaptive training scheme seems to be necessary.

On the contrary, for Innsbruck the estimated effects show a distinct annual
cycle for the location parameters $\mu_\star$, which indicates a varying
information content of the predictor variables $\text{vec}_{\star, mean}$ and
the need of some adaptive training scheme. For the location parameter $\mu_1$,
the intercept is rather large during winter (Fig.~\ref{fig:effects}a)
while, at the same time, the slope coefficient (Fig.~\ref{fig:effects}b)
is close to zero due to an apparently low skill of the EPS. For the location
parameter $\mu_2$ (Fig.~\ref{fig:effects}c,\,d), the higher slope
coefficients during spring and autumn suggest a higher information content of
the raw EPS in the transitional seasons than for the rest of the year. For the
scale parameters (Fig.~\ref{fig:effects}e--h), the estimated effects show
high variability; this indicates a seasonally varying skill of the EPS variance
information. In summary, for the weather station in Innsbruck, the information
content of the ensemble wind predictions seem to be rather low. This is in
accordance with the clearly different pattern of observations and EPS forecasts
shown in Fig.~\ref{fig:climatology}. 

\begin{figure}[!tb] 
\centering 
\includegraphics[width=0.8\textwidth]{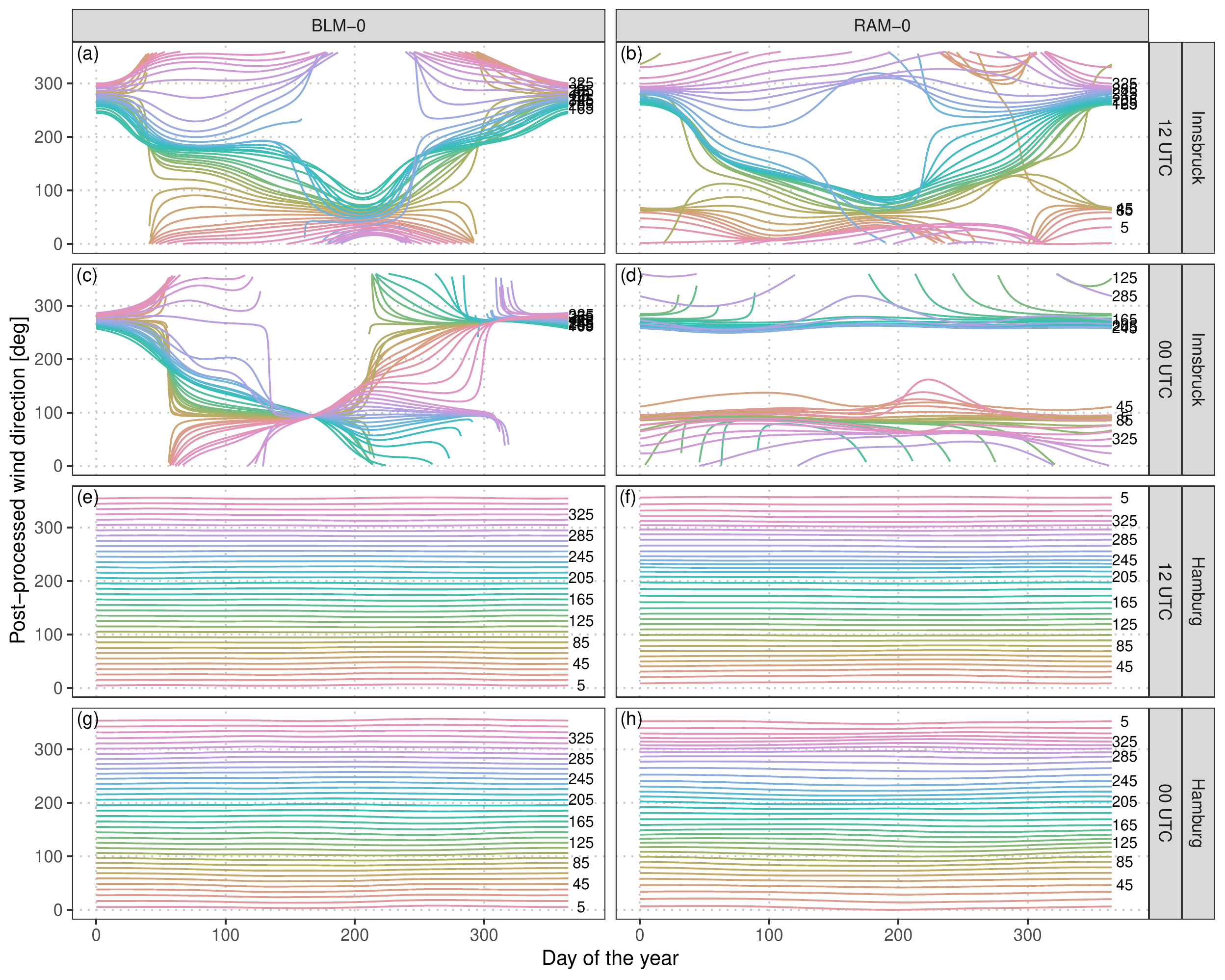}
\caption{Estimated mean effects for the derived post-processed wind direction
at Innsbruck (a--d) and Hamburg (e--h) for the forecast steps $+12$\,h and
$+24$\,h (valid at $12$\,UTC and $00$\,UTC). The colored lines show marginal
effects for the post-processed wind direction conditional on mean values within
10-degree wide wind sectors given the training data set. The effects are
non-centered, and calculated conditional on the day of the year according
to the model setups \mbox{BLM-0} (left, Eq.~(\ref{eq:gam:locsc:v1})) and
 \mbox{RAM-0} (right, Eq.~(\ref{eq:gam:locsc:v2})).}
\label{fig:funcurves} 
\end{figure}

\subsection{Rotation-allowing model without correlation (RAM-0)}\label{sec:results:ram-0}
Figure~\ref{fig:funcurves} shows the estimated mean effects of the setup
\mbox{RAM-0} in comparison with \mbox{BLM-0} on the wind direction at Innsbruck
and Hamburg for the forecast steps $+12$\,h and $+24$\,h valid at $12$\,UTC and
$00$\,UTC, respectively. The marginal effects are non-centered and shown for
the mean covariates within 10 degree wind sectors conditional on the day of the
year. The \mbox{BLM-0} model for Innsbruck shows a distinct seasonal dependency
of the post-processed wind direction for both times of the day
(Fig.~\ref{fig:funcurves}a,\,c). During winter at $12$\,UTC and $00$\,UTC,
mainly down-valley winds (approximately $280$\,degree) are predicted, whereas
over the rest of the year, the EPS mainly forecasts up-valley wind directions.
This pattern is more pronounced during night ($00$\,UTC) and has less
variability in summer. In general, the setup \mbox{BLM-0} seems to mainly
capture the climatological mean wind direction; this leads to little variations
between the different wind directions issued by the EPS. In contrast, the
rotation-allowing setup \mbox{RAM-0} has the flexibility to post-process the
wind directions conditional on the forecasted EPS wind direction, which is
apparent for the station Innsbruck at both $12$\,UTC and $00$\,UTC: For
$12$\,UTC the seasonal dependency leads to either up-valley or down-valley wind
conditional on the ensemble wind direction and on the day of the year
(Fig.~\ref{fig:funcurves}b); whereas at $00$\,UTC almost no seasonal variation
exists and the predicted wind direction solely depends on the issued ensemble
wind forecasts (Fig.~\ref{fig:funcurves}d). At Hamburg, a completely different
picture can be seen: Almost no post-processing conditional on the ensemble wind
direction or the day of the year is visible for both time steps and model
setups (Fig.~\ref{fig:funcurves}e--f). In other words, the predicted ensemble
wind direction fits the observed wind direction quite well and only little
statistical correction is needed.

\subsection{Predictive performance -- models without correlation}\label{sec:results:skill-0}

\begin{figure}[!tb] 
\centering 
\includegraphics[width=0.8\textwidth]{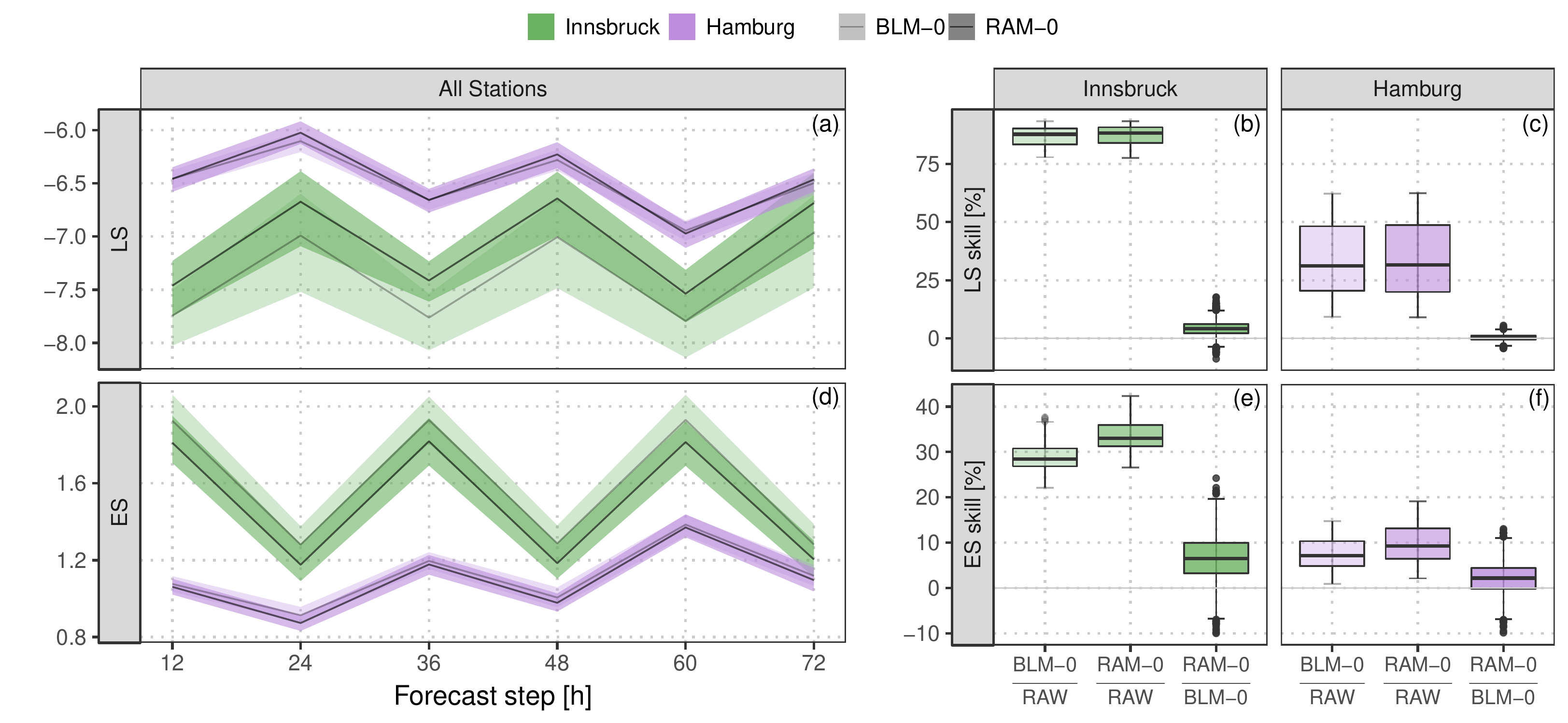}
\caption{Predictive performance in terms of the logarithmic score (LS)
and the energy score (ES) based on the full predictive bivariate distribution
for the out-of-sample validation period. Compared are the two specifications
\mbox{BLM-0} (Eq.~(\ref{eq:gam:locsc:v1})) and \mbox{RAM-0}
(Eq.~(\ref{eq:gam:locsc:v2})). \textbf{Left:} Evolution over time for the
forecast steps from $+12$\,h to $+72$\,h on a $12$~hourly temporal resolution.
The solid lines represent the boot-strapped mean values per forecast step, the
shading the respective $95$\% confidence intervals. \textbf{Right:} Aggregated
skill scores over the forecast steps, comparing the specifications \mbox{BLM-0}
and \mbox{RAM-0} either against the raw ensemble (\mbox{RAW}) or against each
other.} 
\label{fig:locsc:comp} 
\end{figure}

To investigate the predictive performance of the two competing setups,
Fig.~\ref{fig:locsc:comp} shows the discretized logarithmic score (LS; see
Appendix~\ref{app:skillscores}) and the energy score (ES; see
Appendix~\ref{app:skillscores}) for the forecast steps from $+12$\,h to
$+72$\,h on a $12$~hourly temporal resolution. In addition, skill scores are
shown with the raw EPS as reference or comparing the different setups to each
other. Both multivariate scores are proper scores \citep{gneiting.raftery:2007}
and evaluate the full predictive distribution returned by the statistical
models. The scores for the different forecast horizons show an overall better
predictive performance at Hamburg than at Innsbruck. For both stations, the
forecasts valid at $00$\,UTC have more skill than those for $12$\,UTC with
higher diurnal variations at Innsbruck. In terms of the ES, the improvements of
the \mbox{BLM-0} model over the raw EPS is about $29$\% for Innsbruck and $8$\%
for Hamburg (Fig.~\ref{fig:locsc:comp}e,\,f). In terms of the LS
(Fig.~\ref{fig:locsc:comp}b,\,c), the skill scores are higher with an
improvement of approximately $87$\% and $33$\% for Innsbruck and Hamburg,
respectively. The predictive performance gain for the more flexible
rotation-allowing setup \mbox{RAM-0} compared to the specification \mbox{BLM-0}
is around $7$\% for Innsbruck and $2$\% for Hamburg in terms of the ES
(Fig.~\ref{fig:locsc:comp}e,\,f). The LS shows slightly less pronounced
relative improvements for the more flexible setup ($4$\% and $1$\%;
Fig.~\ref{fig:locsc:comp}b,\,c). The distinct improvements in the scores for
\mbox{RAM-0} are as expected for Innsbruck due to a more flexible utilization
of the ensemble information. For plain areas like Hamburg, we assume the better
performance is based on an enhanced adjustment of the location parameters as
both wind components are included in the linear predictors and not due to the
smooth rotation (c.f. Fig.~\ref{fig:funcurves}).
 
\subsection{Rotation-allowing models with correlation}\label{sec:results:ram-x}
After investigating the two competing location or scale setups, we now focus on
an extension of the model \mbox{RAM-0} by explicitly estimating the underlying
correlation structure. Different model specifications for the correlation
parameter $\rho$ are tested employing the same linear predictors for
$\mu_\star$ and $\sigma_\star$ (see Table~\ref{tab:models}).

\begin{figure}[!tb] 
\centering 
\includegraphics[width=0.8\textwidth]{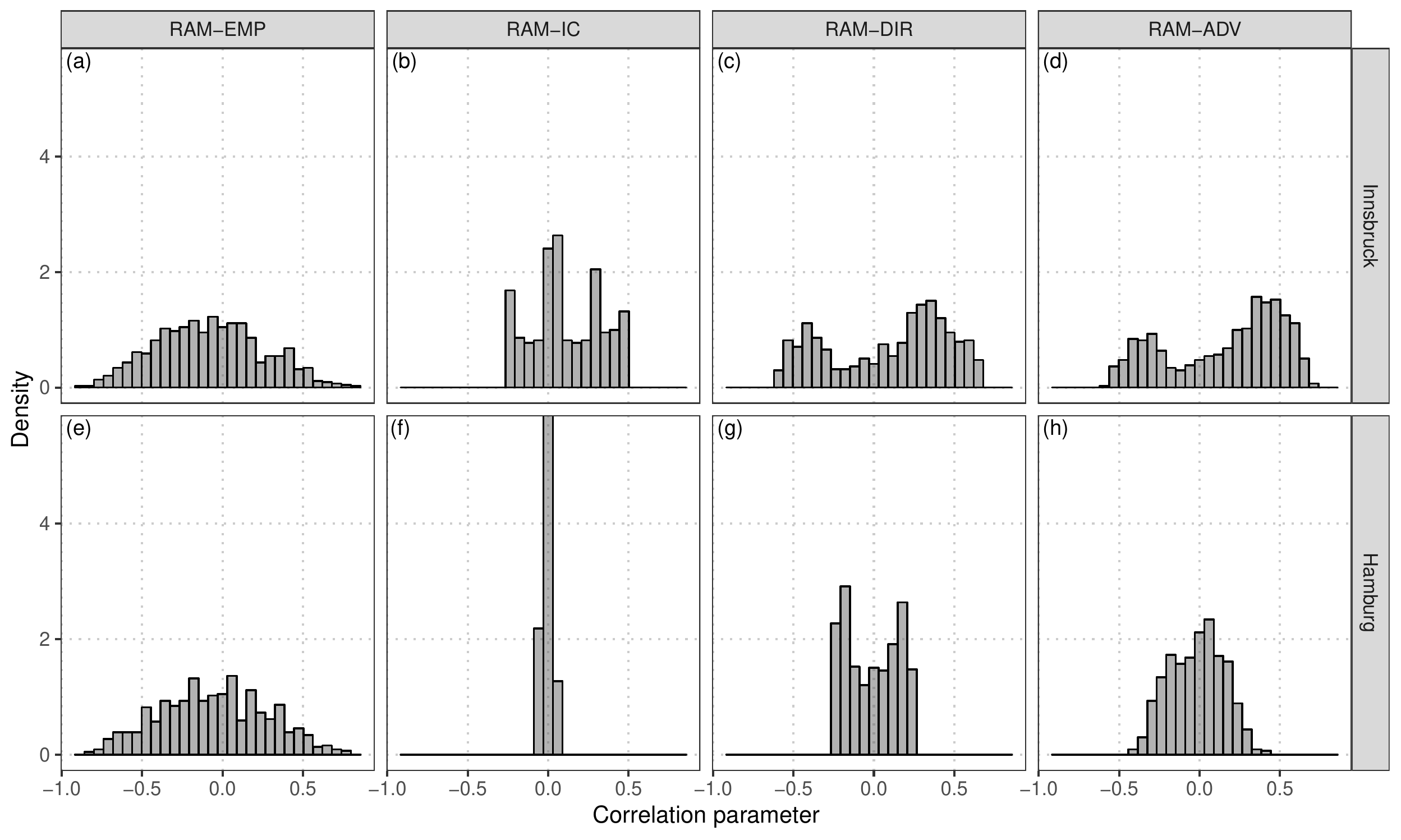}
\caption{Distribution of the correlation parameters for the underlying
dependence structure of the raw ensemble and for the fitted correlation
according to the models specified in Table~\ref{tab:models}. The distributions
are shown for Innsbruck (top row) and Hamburg (bottom row) at the forecast step
$+12$\,h for the out-of-sample validation period.}
\label{fig:cor:dist} 
\end{figure}

Figure~\ref{fig:cor:dist} shows correlation parameters predicted by different
models for the forecast step $+12$\,h for the full validation period. For
comparison, the underlying correlation structure of the raw EPS is also shown.
The latter is distributed similarly for Innsbruck and Hamburg and has almost
the shape of a Gaussian distribution (Fig.~\ref{fig:cor:dist}a,\,e). The
intercept-only model \mbox{RAM-IC}, with a varying intercept over the year,
estimates correlations between $-0.27$ and $0.48$ for Innsbruck
(Fig.~\ref{fig:cor:dist}b) and values near zero without clear seasonal
variations for Hamburg (Fig.~\ref{fig:cor:dist}f). At both stations, the models
with varying effects conditional on the wind direction have similarly
distributed correlation parameters with a slightly larger range of predicted
values for the model \mbox{RAM-ADV} (Fig.~\ref{fig:cor:dist}d,\,f) than for the
setup \mbox{RAM-DIR} (Fig.~\ref{fig:cor:dist}c,\,g). The predicted correlation
parameters are on average larger for Innsbruck than for Hamburg.

\subsection{Predictive performance -- models with correlation}\label{sec:results:skill-x}

\begin{figure}[!tb] 
\centering 
\includegraphics[width=0.7\textwidth]{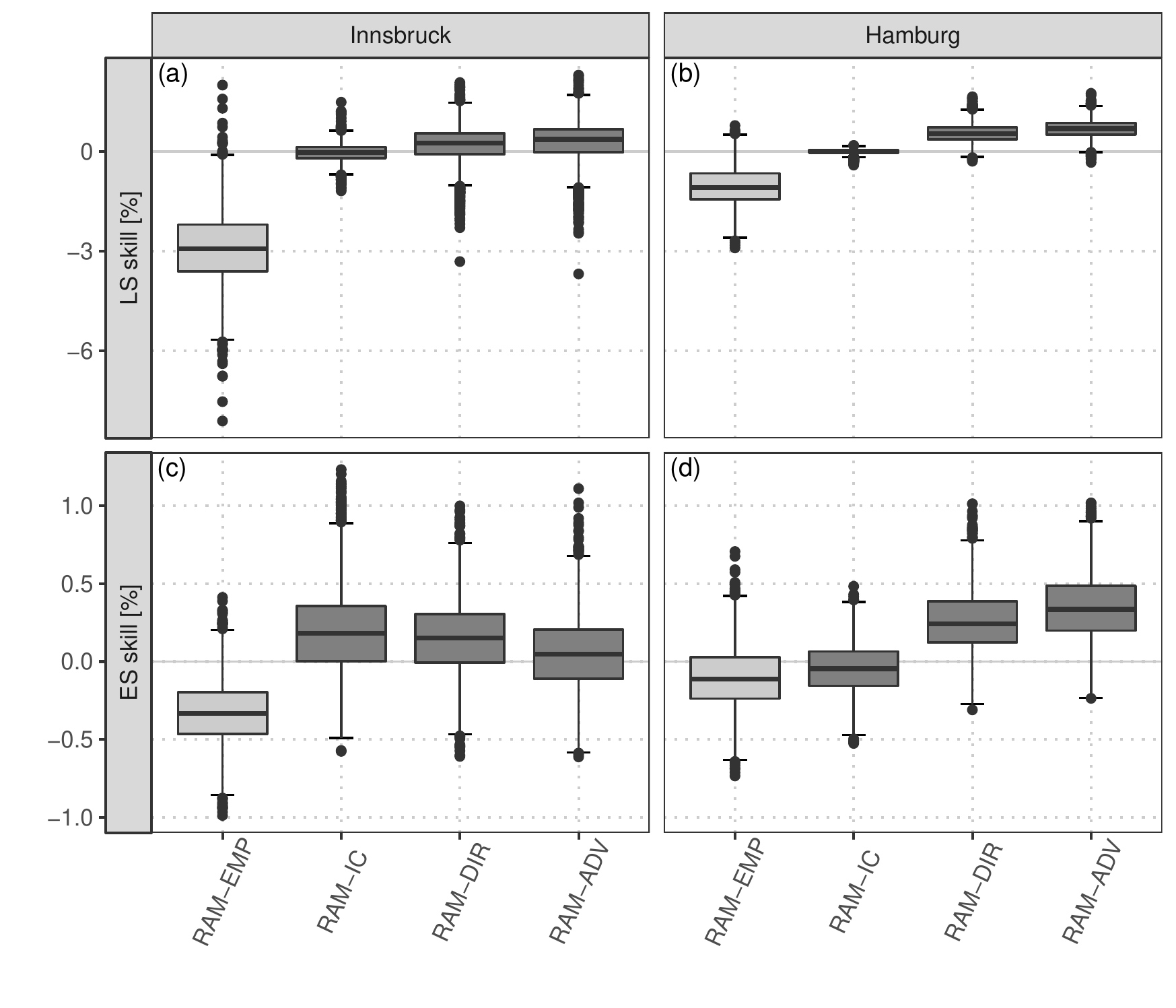}
\caption{Skill scores aggregated over all forecast steps from $+12$\,h to
$+72$\,h on a $12$~hourly temporal resolution based on the full predictive
bivariate distribution for the out-of-sample validation period for Innsbruck
(left) and Hamburg (right). Each box-whisker contains boot-strapped mean values
per forecast step. The scores are shown for the different correlation models
specified in Table~\ref{tab:models} with the univariate post-processed model
assuming a constant correlation of zero (\mbox{RAM-0}) as reference. The
lighter gray color for model \mbox{RAM-EMP} indicates that it uses the
correlation structure of the raw ensemble without further correlation. Skill
scores are in percent, positive values indicate improvements.}
\label{fig:cor:comp} 
\end{figure}

Figure~\ref{fig:cor:comp} shows the verification of bivariate wind speed
predictions with an explicitly estimated correlation parameter for Innsbruck
and Hamburg; the scores are aggregated over the forecast steps $+12$\,h to
$+72$\,h on a $12$~hourly temporal resolution. As in Fig.~\ref{fig:locsc:comp},
the predictive performance is validated in terms of the LS and the ES, based on
the full predictive bivariate distributions. However, for comparing different
predictive distributions with different correlation structures, the ES'
discriminatory ability is limited as it mainly focuses on the location part and
hardly discriminates between different correlation structures
\citep{pinson.tastu:2013}. In Fig.~\ref{fig:cor:comp}, skill scores are shown
for the different correlation models with the post-processed model \mbox{RAM-0}
as reference. The model \mbox{RAM-EMP}, employing the empirical correlation of
the raw EPS, performs slightly worse than the reference model for both stations
and both scores. This indicates that the raw dependence structure of the EPS
has rather low skill. However, for all other models which explicitly model the
correlation only little additional improvement in terms of the LS and the ES is
found. At Innsbruck, the intercept-only model \mbox{RAM-IC} performs best in
terms of the ES (Fig.~\ref{fig:cor:comp}c). Regarding the LS, minor benefits
are present for the most flexible model setup \mbox{RAM-ADV}
(Fig.~\ref{fig:cor:comp}a). For Hamburg, a similar picture is depicted in terms
of the LS (Fig.~\ref{fig:cor:comp}b). For the ES (Fig.~\ref{fig:cor:comp}d),
the model \mbox{RAM-IC} performs slightly worse than the reference model and
the model setup \mbox{RAM-ADV} performs best.

\begin{figure}[!tb] 
\centering 
\includegraphics[width=0.8\textwidth]{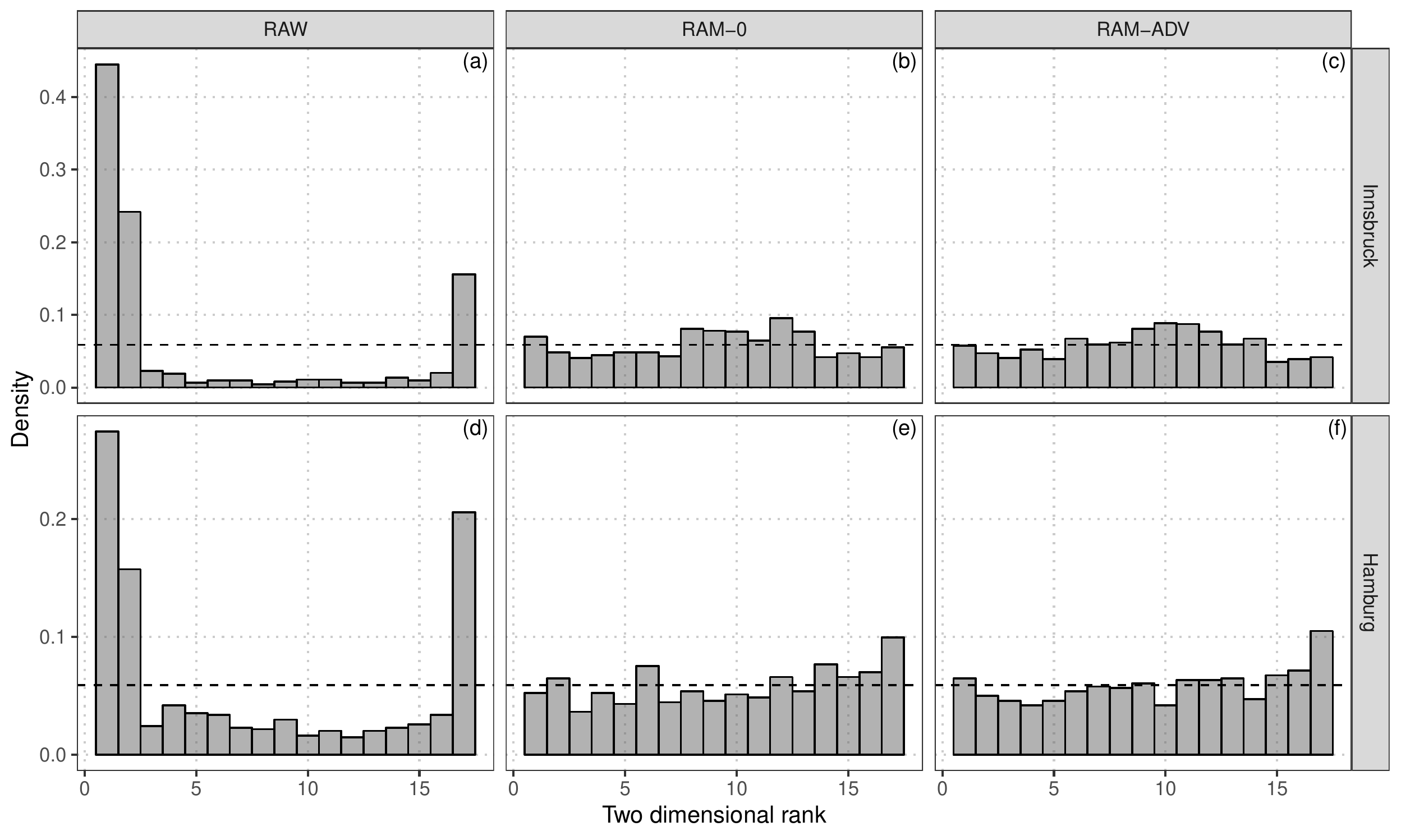}
\caption{Multivariate rank histograms for raw and post-processed ensemble
forecasts according to the correlation model setups \mbox{RAM-0} and
\mbox{RAM-ADV}. The results are shown for Innsbruck (top row) and Hamburg
(bottom row) at the forecast step $+12$\,h for the out-of-sample validation
period. For purposes of presentation, three ranks of the raw EPS are combined
in a single bar. To stabilize the randomness of ties in the calculation of the
multivariate ranks, the median of 20 independent repetitions is plotted.}
\label{fig:pit} 
\end{figure}

To validate the calibration of the post-processed predictions, multivariate
rank histograms \citep{gneiting:2008} are exemplary shown for the model with no
correlation \mbox{RAM-0} and for the model with the most flexible regression
splines in comparison to the raw EPS (Fig.~\ref{fig:pit}). Although the latter
is valid for the grid cell rather than for a single location, both model setups
tested are clearly better calibrated than the highly under-dispersive raw
ensemble. However, for Innsbruck the multivariate rank histograms of the
post-processed forecasts are slightly over-dispersive (Fig.~\ref{fig:pit}b,\,c)
and for Hamburg slightly negatively skewed (Fig.~\ref{fig:pit}e,\,f). The
flexible model setup \mbox{RAM-ADV} shows no significant difference compared to
the model with assumed zero correlation (\mbox{RAM-0}).

\subsection{Evaluation for all sites}\label{sec:results:comparison}

\begin{figure}[!tb] 
\centering 
\includegraphics[width=0.8\textwidth]{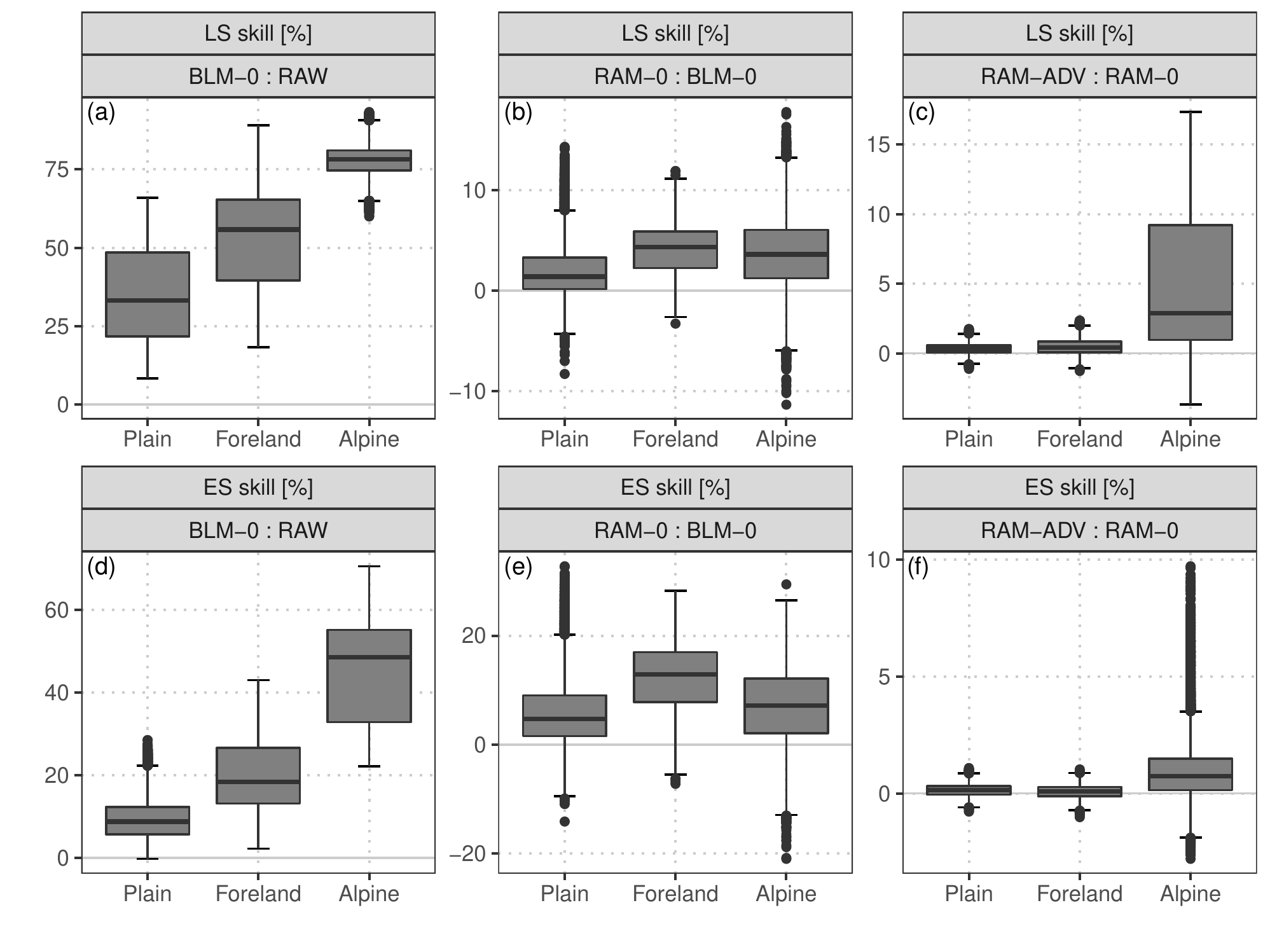}
\caption{Aggregated skill scores (LS: top row, ES: bottom row) for groups of
respective five weather stations which are located in the plain, in the
mountain foreland near the Alps, or within an alpine valley. Each box-whisker
contains boot-strapped mean values of the forecast steps from $+12$\,h to
$+72$\,h on a $12$~hourly temporal resolution for all included stations. The
scores are based on the full predictive bivariate distribution for the
out-of-sample validation period. Compared are the \mbox{BLM-0} model with the
raw EPS as reference, the setup \mbox{RAM-0} with the setup \mbox{BLM-0} as
reference, and the \mbox{RAM-ADV} correlation specification with the
correlation model \mbox{RAM-0} as reference. Skill scores are in percent,
positive values indicate improvements.} 
\label{fig:allscores} 
\end{figure}

After the previous model comparison at two weather stations,
Fig.~\ref{fig:allscores} shows aggregated skill scores for groups of respective
five stations classified as topographically plain, mountain foreland, and
alpine valley sites (see Fig.~\ref{fig:overview}). For the location or scale
models, two comparisons are shown: The \mbox{BLM-0} model is compared to the
raw EPS as reference (Fig.~\ref{fig:allscores}a,\,d), and the more flexible
rotation-allowing setup \mbox{RAM-0} is compared to \mbox{BLM-0}
(Fig.~\ref{fig:allscores}b,\,e). For the correlation specification, the most
flexible model \mbox{RAM-ADV} is compared to the correlation model \mbox{RAM-0}
employing a constant correlation of zero (Fig.~\ref{fig:allscores}c,\,f).

The post-processing employed by the simplest model \mbox{BLM-0} already shows a
distinct improvement over the raw EPS with largest values for alpine valley
sites. In terms of the ES, the skill scores range between mean values of $10$\%
for the plain sites and $45$\% for the alpine valley sites
(Fig.~\ref{fig:allscores}d). A similar picture with an overall larger magnitude
is shown for LS (Fig.~\ref{fig:allscores}a). In the comparison of the two
different setups for the location or scale part
(Fig.~\ref{fig:allscores}b,\,e), the more flexible setup is better regarding
both scores for all station types; largest improvements are found for stations
located in the foreland followed by stations within alpine valleys. The
validation of the correlation models (Fig.~\ref{fig:allscores}c,\,f) shows that
the flexible estimation of the correlation dependence structure is clearly
superior only for station sites within an alpine valley. 

\section{Discussion and conclusion}\label{sec:disc}
In this study, we model the zonal and meridional wind components employing the
bivariate Gaussian distribution in a GAM framework. In contrast to previous
studies all distribution parameters, namely the means and variances for both
wind components but also the correlation coefficient between them, are
simultaneously estimated. The overall performance of the models is evaluated
for three groups of station types classified as topographically plain, mountain
foreland, and alpine valley sites. 

Section~\ref{sec:disc:locs} discusses the benefits of the rotation-allowing
model setup \mbox{RAM-0} over the baseline model BLM-0. In
Sect.~\ref{sec:disc:cor}, the different correlation models are discussed
regarding the potential reason for the low discrimination ability in terms of
predictive performance. At the end, in Sect.~\ref{sec:disc:prop}, a
recommendation is given which statistical model should be used in matter of
simplicity and performance.

\subsection{Rotation-allowing model setup} \label{sec:disc:locs}
The rotation-allowing model (\mbox{RAM-0}) utilizes the ensemble information of
both wind components for both location and both scale parameters in a GAM
framework. This allows the statistical model to adjust for potential
misspecification in the ensemble wind direction by a smooth rotation
conditional on the day of the year and the forecasted wind direction. For
stations in complex terrain, this may be particularly advantageous due to
unresolved topographical features.

The estimated effects confirm a distinct wind rotation for the valley site
(Innsbruck), while for the station in the plain (Hamburg) barely any
adjustments of the forecasted wind direction is needed (see
Fig.~\ref{fig:funcurves}). In terms of predictive performance, the more
flexible model \mbox{RAM-0} outperforms the baseline model \mbox{BLM-0} for
almost all times and stations (see. Fig.~\ref{fig:allscores}b,\,e). However,
the increase in predictive skill is similar for all three station types. This
indicates that -- even if no or only little rotation is needed -- additional
covariates usually yield a better adjustment of the distribution parameters and
therefore an increased predictive skill. Furthermore, the results indicate that
EPS wind forecasts in complex terrain are not solely tilted due to unresolved
valley topographies, but show little skill on average. Thus, for alpine valley
sites the rotation-allowing model mainly captures climatological properties
conditional on the forecasted EPS wind direction. In accordance to this
analysis, larger improvements can be found for stations located in the mountain
foreland where the EPS has a higher information content and a certain rotation
might be necessary.

\subsection{Correlation specifications} \label{sec:disc:cor}
Several different model specifications for the correlation parameter have been
tested. Among others, a flexible setup employing wind direction and speed as
potential covariates for the correlation parameter by non-linear smooth effects
following the idea of \citet{schuhen.etal:2012}. The estimated correlation
parameters seem to be reasonable, and show, on average, larger values for
Innsbruck than for Hamburg (see~Fig.~\ref{fig:cor:dist} for forecast step
$+12$\,h). In terms of predictive skill, all models tested show only minor
improvements compared to the models with zero correlation. The improvements are
highest for stations located inside alpine valleys with a mean improvement of
$1$\% in terms of ES and $5$\% in terms of LS skill scores (see.
Fig.~\ref{fig:allscores}c,\,f). 

\begin{figure}[!tb] 
\centering 
\includegraphics[width=0.8\textwidth]{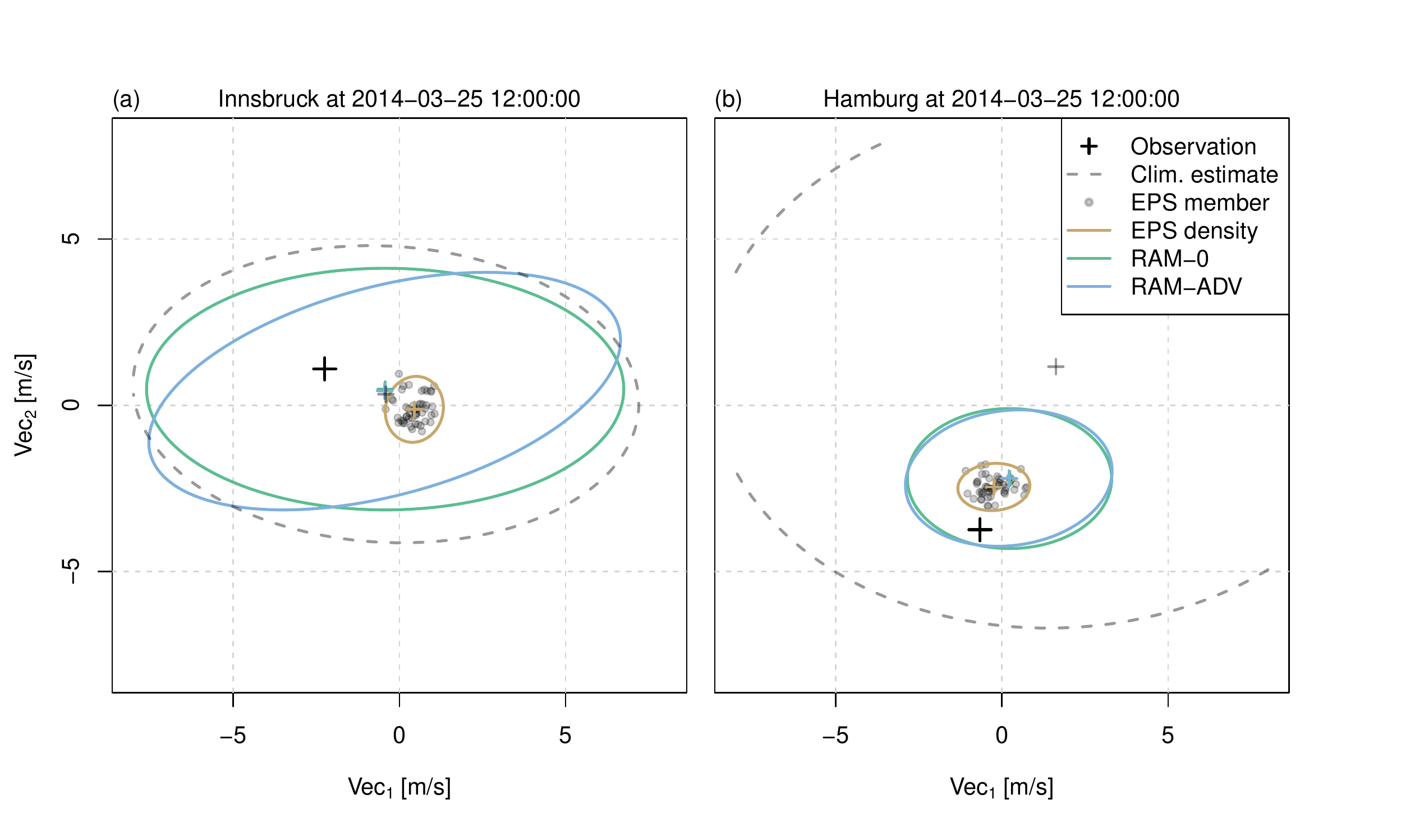}
\caption{Two exemplary forecasts showing the respective observation (black
cross), the climatological estimate (gray dashed line), the EPS member
forecasts (gray points) and their empirical density (brown line), and the
estimated bivariate distributions for the setups \mbox{RAM-0} and
\mbox{RAM-ADV}, without (green line) and with (blue line) modeled correlation,
respectively. The climatological estimate uses the mean, the standard
deviation, and the correlation of the observed wind components as bivariate
distribution parameters. The lines show the $95$\% percentiles of the
respective bivariate distribution; the small crosses the ellipsoid centers or
the location parameters. The shown observations and forecasts are valid at
Innsbruck (left) and Hamburg (right) for March 25, 2014 12\,UTC (forecast
step~$+12$\,h). The forecasts are characteristic for (a)~a valley station and
(b)~a station in the plain within our study.} 
\label{fig:examples} 
\end{figure}

As an illustration of the potential reasons for no more pronounced enhancements
by an explicit estimation of the dependence structure, Fig.~\ref{fig:examples}
shows exemplary forecasts for a station located within an alpine valley
(Innsbruck; Fig.~\ref{fig:examples}a) and in the plains (Hamburg;
Fig.~\ref{fig:examples}b). The figure shows the raw EPS members (gray points)
plus the respective observations (black crosses), climatological estimates
(gray dashed lines), and the corresponding post-processed bivariate
distributions without (green lines) and with (blue lines) an explicitly
estimated correlation parameter. For the valley station, the raw EPS has only
little skill and the uncertainty of the post-processed bivariate distributions
tends towards the climatological estimate. Although a distinct correlation is
estimated by the \mbox{RAM-ADV} model, the variance is still in the same range
as for the \mbox{RAM-0} model. In contrast, for the station in the plain the
uncertainty of the post-processed predictions is much smaller than the
uncertainty of the climatological estimate due to a higher information content
of the EPS. The estimated correlation is close to zero and the predictions of
\mbox{RAM-0} and \mbox{RAM-ADV} look almost identical with a similar elliptic
shape as the raw EPS. This means that for locations where the ensemble provides
only little information, the post-processed uncertainty is rather large and the
statistical model tries to capture unexplained features by the correlation
parameter. For stations where the predictive skill of the raw ensemble is
already high, the statistical models get valuable information about the
expected wind situation and are able to accurately specify the location and
scale parameters. Thus, the correlation of the residuals becomes less important
and typically smaller. This interpretation is supported by the probabilistic
scores used in this study which show improvements in the \mbox{RAM-ADV} models
mainly for alpine valley sites where the skill of the raw ensemble is rather
low.

\subsection{Proposed model specification} \label{sec:disc:prop}
The study shows that the flexible rotation-allowing models bring significant
performance benefits for stations located in complex terrain as well as for
stations in the plain. Therefore, we propose using a similar setup employing
both EPS wind components by a smooth rotation-allowing framework. For
correlation, we have not found a clear distinction between the different
correlation models tested for stations located in the plain and the foreland.
For stations located within an alpine valley, minor improvements could be
found. Despite these somewhat unexpected findings, this has clear advantages
for operational usage: Estimating a single bivariate response distribution
forcing the correlation dependence structure to zero is the same as
post-processing each wind component separately in a univariate setup with
marginal Gaussian response distributions. A univariate post-processing approach
for each respective wind component simplifies the estimation process in terms
of complexity of the required statistical models and reduces computational time
with only little loss of predictive skill at least for the stations tested in
this study.

\clearpage

\section*{Code availability} 
The bivariate Gaussian model estimation is performed in R~3.5.2
\citep{rcoreteam:2018} based on the R package \pkg{bamlss}
\citep{umlauf.etal:2018}. The package provides a flexible toolbox for
distribution regression models in a Bayesian framework. The computation of the
ES is based on the R package \pkg{scoringRules} \citep{jordan.etal:2018}. 

\begin{appendix}
\counterwithin{figure}{section}
\counterwithin{equation}{section}

\section{Model specification complements}
\subsection{Smooth functions}\label{app:gam}
Generalized additive models~\citep[GAM,][]{hastie.tibshirani:1986} and
generalized additive models for location, shape, and
scale~\citep[GAMLSS,][]{rigby.stasinopoulos:2005} are generalizations of linear
regression models which allow one to include potentially non-linear (and even
multi-dimensional) effects in the linear predictors $\eta$. One frequently used
form of non-linear terms are smooth functions, also often referred to as
regression splines. These regression splines are directly linked to the model
parameters as additive terms in $\eta$ and allow the statistical model to
include non-linear transformations of a specific covariate, if needed. For
further details a comprehensive introduction to GAMs is given in
\citet{wood:2017}. An example for an additive predictor $\eta$ with a smooth
function is:
\begin{align} \eta = \alpha_0 +
\underbrace{\vphantom{f_1(\text{x}_{2})}\alpha_1 \cdot
\text{x}_1}_\text{linear effect for x$_1$} +
\underbrace{f_1(\text{x}_{2})}_\text{pot. non-linear effect for x$_2$},
\label{eq:gam:smooth} \end{align}
where $\alpha_\bullet$ are regression coefficients, $\text{x}_\bullet$ the
covariates, and $\alpha_1 \cdot \text{x}_1$ and $f_1(\text{x}_2)$ are a
linear and a potentially non-linear one-dimensional effect, respectively.
Generally, $f_1$ can be any transformation of the covariate $\text{x}_2$
dependent on the specification of $f_1$. For periodic values smooth `cyclic'
splines are often applied, meaning that the function has the same value at its
upper and lower boundaries. This is similar to applying a linear combination of
(several) trigonometric functions, as, e.g., performed by
\citet{schuhen.etal:2012}. In this study, we utilize `cubic' smooth functions
with cyclic constraints \citep{wood:2017}.

\subsection{Time-adaptive training scheme}\label{sec:method:tscheme}
To account for seasonal variations of the intercept and the linear
coefficients, seasonal cyclic splines are used. If the covariates provide
sufficient information a time-adaptive training scheme might not be required.
However, if the bias and/or the slope coefficient is not constant throughout
the year or the covariate's skill varies over the year these terms are
mandatory to allow the statistical model to depict seasonal features.

We therefore fit one statistical model over a training data set including
several years of data but allow the coefficient included in the linear
predictor(s) $\eta$ to smoothly evolve over the year:
\begin{align}
  \eta &= 
  \underbrace{\alpha_0 + f_0(\text{doy})}_{\substack{\text{seasonally varying} \\ \text{intercept}}} ~ + ~ 
  \underbrace{(\alpha_1 + f_1(\text{doy}))}_{\substack{\text{seasonally varying} \\ 
    \text{coefficient for x$_1$}}} \cdot ~ \text{x}_1 
  ~ + ~ \cdots ~ + ~ 
  \underbrace{(\alpha_n + f_n(\text{doy}))}_{\substack{\text{seasonally varying} \\ 
    \text{coefficient for x$_n$}}} \cdot ~ \text{x}_n.
\label{eq:gam:training}
\end{align}
As before, $\alpha_\bullet$ are the regression coefficients, x$_{\bullet}$ the
covariates, and $f_\bullet(\text{doy})$ employ cyclic regression splines
conditional on the day of the year~(doy).

\section{Skill scores used for verification}\label{app:skillscores}
To compare the different bivariate Gaussian models of this study, 
we employ skill scores. A skill score shows the improvements over a
reference. For all measures with a perfect score of zero, the skill score
simplifies to:

\begin{equation}
\text{skill score} = 1 - \frac{\text{score}_\text{forecast}}{\text{score}_\text{reference}}.
\label{eq:skillscores}
\end{equation}

In this study we use the logarithmic score (LS, \citealt{good:1952}) and the
energy score (ES, \citealt{gneiting.raftery:2007}) to validate the probabilistic
performance of the bivariate Gaussian predictions of the statistical
post-processing models. Both multivariate scores evaluate the full predictive
distribution returned by the statistical models. 

The calculation of the ES is based on the R package \pkg{scoringRules}
\citep{jordan.etal:2018}. For a predictive distribution $F$
on $\mathbb{R}^d$ given through $m$ discrete samples
$\mathbf{X}_1,\dots,\mathbf{X}_m$ from $F$ with $\mathbf{X}_i =
(X_i^{(1)},\dots,X_i^{(d)}) \in \mathbb{R}^d, i=1,\dots,m$ the ES can be written as:

\begin{equation}
  \textnormal{ES}(F,\mathbf{y}) = \frac{1}{m}\sum_{i=1}^m \| \mathbf{X}_i - \mathbf{y} \| - 
    \frac{1}{2m^2} \sum_{i = 1}^m\sum_{j = 1}^m \| \mathbf{X}_i - \mathbf{X}_j \|,
\end{equation}

where $\|\cdot\|$ denotes the Euclidean norm on $\mathbb{R}^d$, and $\mathbf{y}
= (y^{(1)},\dots,y^{(d)}) \in \mathbb{R}^d$ the multivariate observation.

The logarithmic score is defined as

\begin{equation}
\textnormal{LS}(F,\mathbf{y}) = \log(f(\mathbf{y})), 
\end{equation}

where the forecast $F$ on $\mathbb{R}^d$ admits a probability density function
$f$, and $\mathbf{y} = (y^{(1)},\dots,y^{(d)}) \in \mathbb{R}^d$ is, as before,
the multivariate observation. For the bivariate Gaussian distribution, the
density function $f$ is identical to the likelihood function $L$ in
Eq.~(\ref{eq:bivnorm}). To be able to calculate skill scores, we compute the LS
by the cumulative distribution function for the neighborhood of the
multivariate observation $\mathbf{y}$, defined by $\mathbf{y} \pm \epsilon$.
Thus, the LS is limited upwards to zero (converges to zero the closer the
forecasts are to perfect predictions), which in turn allows one to calculate
skill scores as given by Eq.~(\ref{eq:skillscores}). In this study, we use
$\epsilon = 0.1$\,ms$^{-1}$; be aware that, depending on $\epsilon$, the
quantitative values of the $LS$ might alter, but the qualitative results remain
unchanged. The calculation of the bivariate cumulative distribution function is
based on the R package \pkg{mvtnorm} \citep{genz.bretz:2009}. 

\end{appendix}

\section*{Acknowledgements}
This project was funded by the Austrian Research Promotion Agency (FFG), grant
no. 858537. We also thank the Zentralanstalt f\"ur Meteorologie und Geodynamik
(ZAMG) for providing access to the data. 

\clearpage

\bibliography{analysis_bivnorm}

\end{document}